\documentclass[12pt]{article}
\usepackage{setspace}
\usepackage[english]{babel}
\usepackage{pdfpages}
\usepackage{multicol}
\usepackage[letterpaper,top=2cm,bottom=2cm,left=1in,right=1in,marginparwidth=1in]{geometry}

\usepackage{subcaption}
\usepackage[colorlinks=true, allcolors=blue]{hyperref}

\usepackage[comma]{natbib}%[square] super,sort&compress,
\usepackage{makecell}
\usepackage{lscape}
\usepackage{caption}
\usepackage{bm}

\usepackage{docmute}

\definecolor{blue}{rgb}{0,0,1}

\newcommand\locrone[1]{}%{\printTodo{#1}}%
\newcommand{\revone}{}%\color{blue}}%

\usepackage{amsmath}

\usepackage{xr}

\usepackage[utf8]{inputenc}
\usepackage[T1]{fontenc}
\usepackage{hyperref}
\usepackage{url}
\usepackage{booktabs}
\usepackage{array}
\usepackage{tabularx}
\usepackage{longtable}
\usepackage{multirow}
\usepackage{amsfonts}
\usepackage{amsmath,amssymb}
\usepackage{nicefrac}
\usepackage{microtype}
\usepackage{xcolor}
\usepackage{graphicx}
\usepackage{tikz}
\usetikzlibrary{arrows.meta,positioning,calc}

\usepackage{color}
\definecolor{blue}{rgb}{0,0,1}

\newcommand{\prove}{\textsc{PROVE}}

\newcommand{\artifactplaceholder}[1]{\textcolor{blue}{[ARTIFACT PLACEHOLDER: #1]}}

\IfFileExists{generated/rule_set_a_results_macros.tex}{% Generated by experiments/run_rule_set_a_simulation.py

}{}

\IfFileExists{generated/rule_set_a_v5_results_macros.tex}{% Generated by experiments/run_rule_set_a_simulation.py

}{}

\IfFileExists{generated/rule_set_a_v1_results_macros.tex}{% Generated by experiments/run_rule_set_a_simulation.py

}{}
\newcolumntype{Y}{>{\raggedright\arraybackslash}X}

\title{Bridging Probabilistic LLMs and Deterministic Statistical Validation: The PROVE Multi-Agent Framework for Clinical Trial Reporting}

\author{%
  Zhaohua Lu\textsuperscript{*}, Cheng Zheng\textsuperscript{*}, Yuanyuan Han\textsuperscript{\$}\\
  \textsuperscript{*} Biostatistics and Data Management, Daiichi-Sankyo Inc. \\  
  \textsuperscript{\$} Data and Quantitative Sciences, Bristol Myers Squibb (BMS) %\\
}
\date{June 12, 2026}

\makeatletter

\newcommand*{\addFileDependency}[1]{% argument=file name and extension
\typeout{(#1)}% latexmk will find this if $recorder=0
\@addtofilelist{#1}
\IfFileExists{#1}{}{\typeout{No file #1.}}
}\makeatother

\begin{document}

\maketitle

\begin{abstract}
Ensuring the accuracy and consistency of clinical trial Tables, Figures, and Listings (TFLs) remains a central challenge in regulatory reporting. Independent programming and manual review are essential quality-control practices, but cross-output verification still relies heavily on reviewer inspection and can miss structural, logical, or arithmetic discrepancies. Large language models (LLMs) may help interpret varied table language and navigate long study documents, but they are not reliable substitutes for programmed statistical checks. 
We introduce \prove{} (Programmatic Reporting and Output Verification Engine), an auditable framework that uses optional LLM and retrieval support for table interpretation while reserving numerical and logical decisions for programmed validators. \prove{} links findings to source evidence, supports cross-output consistency checks, and allows LLM use to be enabled or disabled according to study requirements. We evaluated \prove{} using ten replicated synthetic oncology reporting packages generated from raw data through SDTM, ADaM, and TFLs outputs, with paired clean and discrepancy-injected packages; each replicate included 15 randomly injected discrepancies. Two table-label settings were studied: exact labels matching the validator vocabulary and label variations with similar clinical meaning but different wording.
{\revone Within the implemented rule classes, all automated \prove{} variants achieved perfect classification on clean and discrepancy-injected packages in the exact-label setting.\locrone{R1.1.1}} In the label-variation setting, LLM-assisted semantic matching improved overall recall from 0.588 to 0.993 and overall F1 from 0.735 to 0.996 compared with the exact-match, fuzzy lexical, and embedding similarity variants. These findings suggest that LLMs are most useful for robust interpretation of real-world variation in TFL wording and formatting, while executable checks should remain responsible for final numerical validation.
\end{abstract}

\section{Introduction}

In drug development, clinical trial reporting necessitates the generation of an extensive collection of statistical outputs to serve as evidence for decision-makers, including regulatory authorities and study sponsors. To ensure these outputs remain mutually consistent and accurate across source data, analysis datasets, statistical analysis plans (SAPs), protocols, and clinical study report materials, sponsors traditionally utilize independent programming, manual review, and visual quality control to identify discrepancies in TFLs. While these established practices remain essential and effective, maintaining total output quality continues to be a significant challenge. For instance, late-stage errata in oncology advisory committee materials illustrate that arithmetic, censoring, denominator, and subgroup discrepancies can still propagate into final regulatory-facing documents.\citep{eflornithine2023,sotorasib2023,imetelstat2024,carvykti2024,imfinzi2024}.

The increasing use of AI in the industry has sharpened the need for transparent, context-specific, and risk-based validation. FDA guidance emphasizes credibility assessment for AI models in their context of use \citep{fda2025_ai_guidance}, while FDA/EMA Good AI Practice principles emphasize human-centric design, clear context of use, documentation, performance assessment, and lifecycle management \citep{fda_ema_2026_good_ai}. Reporting guidelines such as CONSORT-AI, SPIRIT-AI, and DECIDE-AI similarly stress transparent description of AI components, data flow, user interaction, and evaluation methodologies \citep{liu2020consortai,cruzrivera2020spiritai,vasey2022decideai}.

%This paper studies a high-impact trustworthy-AI and practical problem: This paper addresses the critical intersection of trustworthy AI and regulatory reporting by exploring the role of LLMs in automating quality control for clinical trial outputs. Specifically, we investigate a multi-agent framework designed to assist in deterministic statistical verification, ensuring that probabilistic language models enhance semantic interpretation without compromising the mathematical integrity required for drug development reporting. We present \prove{} (Programmatic Reporting and Output Verification Engine), a role-separated LLM-orchestrated system in which probabilistic components support extraction, semantic interpretation, routing, and optional reference grounding, while executable validators perform arithmetic and logical checks.

This research addresses the intersection of trustworthy AI and regulatory reporting by exploring how LLMs can assist quality control for clinical trial outputs. We investigate a multi-agent framework in which LLMs help interpret document content, but do not decide whether reported numbers are correct. We introduce \prove{} (Programmatic Reporting and Output Verification Engine), a system where LLM components help with extraction and document lookup, while independent executable validators perform arithmetic and logical checks.

\paragraph{Contributions.}
%First, we provide an automated and scalable framework, the \prove{} architecture, which improves reporting precision and checking efficiency by separating LLM/RAG-assisted interpretation from deterministic statistical validation. Second, we introduce a semantic reporting graph that maps table cells and extracted claims into provenance-linked clinical and statistical concepts to enable robust cross-output verification. Third, we implement operational controls for binary LLM/API usage, evidence-audit tracking, and data provenance to ensure transparency and accountability in the review process. Fourth, we introduce a labeled rule-violation benchmark and a comprehensive simulation study to rigorously quantify true positives, false positives, precision, recall, F1-score, and rule-mapping behavior.
We make four primary contributions to AI-assisted clinical reporting. First, we provide the \prove{} scalable architecture, which improves checking efficiency by using LLMs for interpretation while reserving mathematical verification for deterministic statistical validators. Second, we introduce a semantic reporting graph that connects table cells and extracted statements to clinical concepts, source locations, and related outputs. Third, we implement practical review safeguards: findings can be linked back to evidence, and reviewer-facing outputs record how each finding was produced. Fourth, we present a reproducible simulation study that generates complete clinical reporting packages and evaluates labeled discrepancies using precision, recall, and rule-mapping accuracy.

\section{The \prove{} Framework}

%\prove{} ingests clinical reporting packages containing structured files, tables, listings, figures, optional protocol/SAP documents, and optional user-defined rules. The system normalizes populations, arms, endpoints, visits, and row/column semantics; extracts output claims; links claims to source artifacts; and runs deterministic validators over structured representations. LLM/RAG components can be used where semantic interpretation or document retrieval is required.
\prove{} processes clinical reporting packages by ingesting historical and current TFL outputs and study documents such as protocols or SAPs. The framework standardizes trial populations, arms, endpoints, and visits so that related outputs can be compared consistently. Reported values and statements are linked back to their source files, so reviewers can trace each finding to the table, listing, or document that produced it. To maintain statistical integrity, deterministic validators execute all logical and mathematical checks. Within this workflow, LLM and retrieval components are used only for language interpretation and document search; final verification remains grounded in reproducible code.

\begin{figure}[t]
\centering
\resizebox{\linewidth}{!}{%
\begin{tikzpicture}[
  node distance=0.9cm and 0.9cm,
  box/.style={draw, rounded corners=1pt, align=center, minimum width=2.45cm, minimum height=0.9cm, font=\large},
  graphbox/.style={draw, rounded corners=1pt, align=center, minimum width=3.2cm, minimum height=1.2cm, fill=gray!12, font=\large\bfseries},
  sidebox/.style={draw, rounded corners=1pt, align=center, minimum width=2.7cm, minimum height=0.9cm, fill=blue!6, font=\large},
  outbox/.style={draw, rounded corners=1pt, align=center, minimum width=2.7cm, minimum height=0.9cm, fill=green!6, font=\large},
  arrow/.style={-{Latex[length=2.2mm]}, thick},
  sidearrow/.style={-{Latex[length=2.2mm]}, thick, dashed},
  legendlabel/.style={font=\normalsize, anchor=west}
]
\node[box] (input) {Clinical reporting\\package};
\node[box, right=of input] (extract) {Ingestion and\\extraction};
\node[box, right=of extract] (normalize) {Standardization\\of terms and values};
\node[graphbox, right=of normalize] (graph) {Semantic reporting graph\\\normalfont concepts, denominators,\\hierarchies, versions, source links};
\node[box, right=of graph] (validators) {Deterministic validators\\and user rules};
\node[box, right=of validators] (reconcile) {Combine findings\\and apply settings};
\node[outbox, right=of reconcile] (outputs) {Review queue, audit\\report, metrics};

\node[sidebox, above=1.35cm of normalize] (rag) {SAP/protocol\\retrieval};
\node[sidebox, right=of rag] (cross) {Cross-output checks\\concept, denominator,\\source-link, version checks};
\node[sidebox, right=of cross] (adjudication) {Evidence summaries\\for review};

\draw[arrow] (input) -- (extract);
\draw[arrow] (extract) -- (normalize);
\draw[arrow] (normalize) -- (graph);
\draw[arrow] (graph) -- (validators);
\draw[arrow] (validators) -- (reconcile);
\draw[arrow] (reconcile) -- (outputs);

\draw[sidearrow] (rag) to[bend left=10] (graph);
\draw[sidearrow] (graph) -- (cross);
\draw[sidearrow] (cross) to[bend right=12] (validators);
\draw[sidearrow] (graph) -- (adjudication);
\draw[sidearrow] (adjudication) -- (outputs);

\begin{scope}[shift={($(outputs.north east)+(-5.8cm,3.0cm)$)}]
  \node[draw, rounded corners=1pt, anchor=north west, inner sep=3pt, minimum width=5.8cm, minimum height=0.85cm] at (0,0) {};
  \node[legendlabel] at (0.15,-0.22) {Legend};
  \draw[arrow] (1.15,-0.22) -- (1.95,-0.22);
  \node[legendlabel] at (2.10,-0.22) {primary path};
  \draw[sidearrow] (1.15,-0.58) -- (1.95,-0.58);
  \node[legendlabel] at (2.10,-0.58) {supporting flow};
\end{scope}
\end{tikzpicture}}
\caption{\prove{} architecture and data flow. Legend: solid arrows show the primary processing path from input package to reviewer-facing outputs, and dashed arrows show supporting information flows such as SAP/protocol retrieval, cross-output checks, and evidence summaries. The semantic reporting graph is the shared representation between extraction/standardization and programmed validation.}
\label{fig:architecture}
\end{figure}
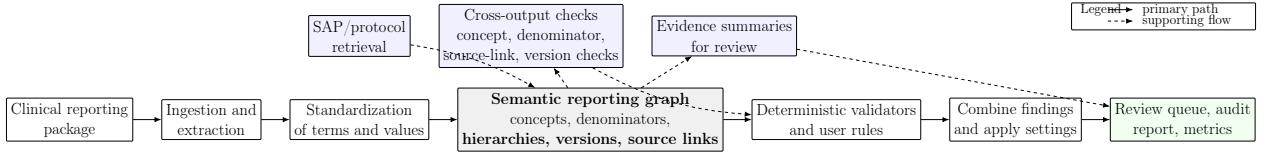

\subsection{Semantic Reporting Graph for Cross-Output Verification}

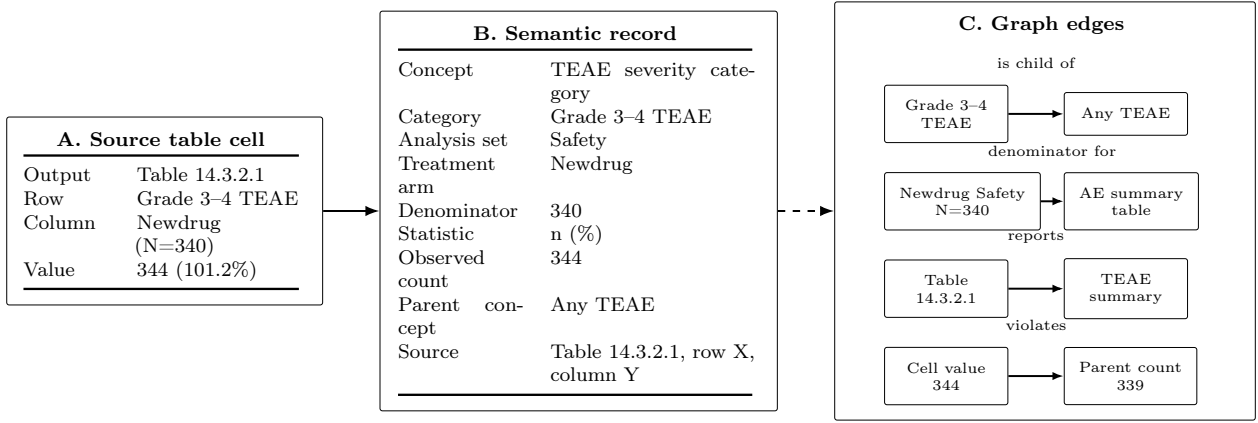
\begin{figure}[ht]
	\centering
	\resizebox{\linewidth}{!}{%
		\begin{tikzpicture}[
			panel/.style={draw, rounded corners=1pt, inner sep=6pt, align=left},
			arrow/.style={-{Latex[length=2.0mm]}, thick},
			link/.style={-{Latex[length=2.0mm]}, thick, dashed},
			font=\scriptsize
			]
			\node[panel] (source) {
				\begin{minipage}{0.24\linewidth}
					\centering
					\scriptsize
					\textbf{A. Source table cell}\\[0.3em]
					\begin{tabular}{@{}p{0.30\linewidth}p{0.58\linewidth}@{}}
						\toprule
						Output & Table 14.3.2.1 \\
						Row & Grade 3--4 TEAE \\
						Column & Newdrug (N=340) \\
						Value & 344 (101.2\%) \\
						\bottomrule
					\end{tabular}
				\end{minipage}
			};
			
			\node[panel, right=0.8cm of source] (knot) {
				\begin{minipage}{0.31\linewidth}
					\centering
					\scriptsize
					\textbf{B. Semantic record}\\[0.3em]
					\begin{tabular}{@{}p{0.34\linewidth}p{0.56\linewidth}@{}}
						\toprule
						Concept & TEAE severity category \\
						Category & Grade 3--4 TEAE \\
						Analysis set & Safety \\
						Treatment arm & Newdrug \\
						Denominator & 340 \\
						Statistic & n (\%) \\
						Observed count & 344 \\
						Parent concept & Any TEAE \\
						Source & Table 14.3.2.1, row X, column Y \\
						\bottomrule
					\end{tabular}
				\end{minipage}
			};
			
			\node[panel, right=0.8cm of knot] (graphpanel) {
				\begin{minipage}{0.33\linewidth}
					\centering
					\scriptsize
					\textbf{C. Graph edges}\\[0.3em]
					\begin{tikzpicture}[
						x=1cm,
						y=1cm,
						gnode/.style={
							draw,
							rounded corners=1pt,
							align=center,
							minimum width=1.75cm,
							minimum height=0.55cm,
							font=\tiny
						},
						garrow/.style={-{Latex[length=1.6mm]}, thick},
						edgelabel/.style={
							font=\tiny,
							anchor=south
						}
						]
						\node[gnode] (grade)  at (0, 0)    {Grade 3--4\\TEAE};
						\node[gnode] (any)    at (2.55, 0) {Any TEAE};
						
						\node[gnode] (denom)  at (0, -1.24)    {Newdrug Safety\\N=340};
						\node[gnode] (ae)     at (2.55, -1.24) {AE summary\\table};
						
						\node[gnode] (table)  at (0, -2.48)    {Table\\14.3.2.1};
						\node[gnode] (summary) at (2.55, -2.48) {TEAE\\summary};
						
						\node[gnode] (value)  at (0, -3.72)    {Cell value\\344};
						\node[gnode] (parent) at (2.55, -3.72) {Parent count\\339};
						
						\draw[garrow] (grade) -- (any);
						\draw[garrow] (denom) -- (ae);
						\draw[garrow] (table) -- (summary);
						\draw[garrow] (value) -- (parent);
						\node[edgelabel] at ($(grade.east)!0.5!(any.west)+(0,0.43)$) {is child of};
						\node[edgelabel] at ($(denom.east)!0.5!(ae.west)+(0,0.43)$) {denominator for};
						\node[edgelabel] at ($(table.east)!0.5!(summary.west)+(0,0.43)$) {reports};
						\node[edgelabel] at ($(value.east)!0.5!(parent.west)+(0,0.43)$) {violates};
					\end{tikzpicture}
				\end{minipage}
			};
			
			\draw[arrow] (source) -- (knot);
			\draw[link] (knot) -- (graphpanel);
	\end{tikzpicture}}
	\caption{Example semantic-record construction. A source TFL cell is standardized into a source-linked semantic record, then connected to graph edges that support programmed cross-output checks. Solid arrows indicate direct transformation of extracted content; the dashed arrow indicates linking the semantic record into the broader reporting graph.}
	\label{fig:semantic_knot_example}
\end{figure}

A fundamental design choice in \prove{} is to transform extracted table cells, row statements, and listing-derived quantities into a shared semantic reporting graph. The graph links reported values to their clinical meaning, source context, and related outputs, allowing programmed validators to enforce cross-output relationships such as safety hierarchy and denominator consistency. Figure~\ref{fig:semantic_knot_example} illustrates this transformation from a source table cell to a source-linked semantic record and graph edges.

%Graph construction utilizes a hybrid methodology. LLM and RAG components are restricted to semantic tasks that require language interpretation or document context. These tasks include proposing concept labels, linking row labels to protocol-defined endpoints, and retrieving supporting reference passages from the SAP. For instance, the LLM can autonomously map non-standard row descriptors in a draft table to their corresponding CDISC-compliant nodes in the graph based on the SAP’s definitions. For example, an LLM can reconcile semantically similar but syntactically different terms across documents, for example, mapping 'Treatment-Emergent Signs and Symptoms' in a legacy SAP to the 'TEAE' node in a standardized graph. While a deterministic parser might fail to link these due to the string mismatch, the LLM identifies the conceptual equivalence, allowing the system to then programmatically verify that the underlying patient counts remain consistent across the disparate outputs. These components produce candidate semantic anchors with full provenance but do not contribute to numerical computation or final decision-making.

Graph construction is role separated. Programmed logic handles reproducible structure, denominator, relationship, and graph-construction steps, while optional LLM and retrieval components are limited to language tasks such as proposing concept labels, aligning row descriptions with protocol endpoints, and retrieving SAP context. These probabilistic mappings may help identify what a reported value means, but they do not perform numerical computation or make final correctness decisions, and they can be disabled according to study requirements. Because semantic records and graph links retain source references, each finding remains traceable to the corresponding output cell and supporting evidence, and the final validation result remains reproducible from the graph data and rule configuration.

\subsection{Deterministic validators and rule support}

The deterministic validation layer is organized around the kinds of consistency questions that arise in clinical trial reporting. Basic summary checks cover counts, percentages, denominators, and categorical totals. Statistical checks cover descriptive summaries, confidence intervals, p-values, and treatment-effect displays. Endpoint-specific checks address time-to-event outputs such as Kaplan--Meier summaries, while safety checks evaluate adverse-event hierarchies and related count constraints. Cross-output checks then compare related tables, listings, figures, and prior versions so that a finding is not limited to the table in which it first appears.

Rules enter the framework in three complementary ways. Built-in validators encode recurring clinical reporting checks that can be run directly on the extracted outputs. Sponsor-defined executable rules allow a study team to express additional review requirements using controlled rule templates, which are then mapped to local deterministic checks. LLM-backed natural-language rules are treated separately: they can retrieve SAP or protocol context and propose review items, but deterministic validators or human reviewers remain responsible for deciding whether an output is incorrect. Table \ref{tab:rule_examples} illustrates these rule types and their execution roles.

\begin{table}[ht]
	\caption{Illustrative rule types supported by \prove{}.}
	\label{tab:rule_examples}
	\centering
	\begin{tabularx}{\linewidth}{>{\raggedright\arraybackslash}p{0.22\linewidth}YY}
		\toprule
		Rule type & Example expression & Execution role \\
		\midrule
		Internal deterministic validator & For an $n$ (\%) cell with denominator $N$, recompute $100n/N$ and compare with the displayed percentage according to specified rounding rules. & Executed in Python over extracted table cells; reports expected and observed values, tolerance, and source row/column. \\ \hline
		Cross-output consistency check & A safety summary table, related AE table, and supporting listing should use compatible safety-population denominators and event counts. & Compares related outputs after extraction and standardization; reports the affected outputs and the source evidence used for review. \\ \hline
		Sponsor-defined executable rule & A rule entry mapping the label ``Child count cannot exceed parent'' to a hierarchy-check key with associated table and row parameters. & Maps domain-specific parameters to a local executable validator and records the sponsor rule identifier for review. \\ \hline
		LLM-backed semantic rule & ``Using SAP Section 6.4, verify that PFS censoring footnotes and event definitions in the TFLs follow the planned analysis.'' & Retrieves SAP passages to propose labels or review items; deterministic validators or human review decide whether an output violates the grounded rule. \\
		\bottomrule
	\end{tabularx}
\end{table}

\subsection{Review Controls and Auditability}

%\prove{} exposes two simple run-time safeguards. First, LLM control is binary: \texttt{off} prevents API-backed calls, while \texttt{on} requires an available API key and uses a retry allowance for transient API failures.

%Each finding carries structured audit fields, including rule identifier, error type, expected and observed values, table/row/column location, source pages when available, deterministic signature, origin task, evidence citations, evidence-audit status, and review-queue state. This metadata allows a reviewer to distinguish a reproducible arithmetic finding from a document-grounded semantic finding, trace the finding back to source evidence, and rerun the same setting for verification.

\prove{} is designed to return findings in a form that supports reviewer action, not only algorithmic detection. For each potential discrepancy, the engine reports what was checked, what value was observed, what value or relationship was expected, and where the issue appears in the reporting package. When supporting material is available, the finding is connected to relevant evidence such as the source table cell, listing record, SAP or protocol passage, or corresponding value in a prior output version.

This evidence-centered output helps users triage findings more efficiently. A reviewer can distinguish a direct arithmetic discrepancy from an interpretation that depends on SAP wording, confirm whether the reported evidence supports the finding, and follow the links back to the original source material. The same structure also supports accountability: findings are connected to the deterministic check or review rule that produced them, and optional LLM-assisted evidence retrieval is presented as supporting context rather than as the final decision-maker.

\section{Simulation Study}

We simulate a synthetic Phase 3 oncology trial for previously treated non-small cell lung cancer. For each replicate, the data-generation pipeline creates raw subject-level source data, derives SDTM domains, derives ADaM analysis datasets, and generates a full TFL package. The simulation includes disposition, demographics and baseline disease characteristics, prior and concomitant therapies, treatment exposure, adverse events, progression-free survival, tumor response, and duration of response. Table~S1 summarizes the generated content used in the simulation.
%\ref{tab:generated_content} 

The numerical experiment uses a paired clean/discrepancy-injected design. For each replicated reporting package, the clean TFL package is preserved and a paired package is created by injecting 15 randomly selected discrepancies into the same outputs. This design allows the evaluation to measure both detection of known discrepancies and false alarms when the corresponding reporting package is clean. The discrepancies are drawn from two clinically motivated groups: direct inconsistencies within full TFL outputs and inconsistencies that require comparison across related outputs or listings. In broad terms, the rules cover percentage and denominator checks, descriptive-statistic consistency, categorical totals, confidence-interval and p-value logic, time-to-event accounting, adverse-event hierarchy checks, exposure-threshold checks, and cross-output consistency. Rule descriptions are summarized in Table~S2. {\revone These discrepancy types were selected for targeted evaluation rather than to estimate the full natural distribution of clinical reporting discrepancies. The benchmark therefore measures detection performance for discrepancies sampled from the implemented rule classes.
\locrone{R1.1.2}}
%\ref{tab:rule_descriptions}

We evaluate two table-label settings. In the first setting, row, column, and statistic labels agree with the predefined labels used by the validator engine. This version measures performance when the reporting package follows the expected terminology. In the second setting, selected row, column, and statistic labels are deliberately changed to new wording with similar clinical or statistical meaning. This label-variation version creates a more realistic challenge: the underlying numbers are unchanged in structure, but the method must still recognize that labels such as alternative response, safety, or time-to-event descriptions refer to the same reporting concepts. {\revone Examples of exact-label and label-variation wording are provided in Supplementary Table~S3.%\ref{tab:label_variation_examples}.}

%The comparison uses four automated \prove{} variants, summarized in Table~\ref{tab:evaluation_settings}. 
%The non-LLM variants use exact rule-label matching, fuzzy lexical matching, or embedding similarity matching for label interpretation while disabling LLM-assisted interpretation. 

{\revone The comparison uses four automated \prove{} variants and one manual-review approach. The automated variants share the same programmed numerical validators and differ only in the strategy used to map observed table labels to canonical semantic labels before validation: exact rule-label matching, fuzzy lexical matching, embedding similarity matching, and LLM-assisted semantic matching. The exact-match variant requires labels to match the validator vocabulary. The fuzzy variant uses Python's \texttt{difflib.SequenceMatcher} to compare normalized observed labels with candidate semantic labels, using a default match threshold of 0.85 and a required 0.05 margin over the runner-up match. The embedding variant uses \texttt{sentence-transformers/all-MiniLM-L6-v2} through the \texttt{sentence-transformers} package and scores label embeddings by cosine similarity, using a default threshold of 0.74 and the same 0.05 runner-up margin. 
These variants represent environments where LLM use is restricted by policy, infrastructure, or validation requirements, and they provide the non-LLM reference conditions for the study.
%The LLM-assisted variant uses language-model support for table interpretation and related semantic mapping tasks while retaining programmed validators for all numerical and logical decisions.
}

The LLM-assisted \prove{} variant enables LLM assistance for table interpretation and related language tasks, such as recognizing variant row labels or mapping semantically similar output terms, while retaining programmed validators for all numerical and logical decisions. This variant evaluates whether language-model support improves detection in less standardized reporting packages without transferring final correctness decisions to the LLM. 
{\revone In this simulation, LLM-assisted PROVE is supported by ChatGPT 5.2 API.} 
The comparison therefore focuses on the incremental contribution of LLM-assisted interpretation within the same programmed validation framework.

{\revone
Manual review was included as an independent human-review baseline. Seven simulation replications were randomly selected from the 10 simulated replications used by PROVE, and each replication was assigned to one independent biostatistician. Reviewers were asked to inspect the oncology topline output batch (SDTM datasets, ADaM datasets and the Word outputs) without using AI tools, although programming checks in R or SAS were allowed. The review instruction was to focus primarily on table-body content rather than formatting. Reviewers marked suspected issues directly in the Word tables rather than maintaining a separate issue log. The review lasted for at least 15 minutes, and continued review for up to another 45 minutes was encouraged, roughly more than twice the observed runtime of the LLM-assisted \prove{} variant. 
}

%\begin{table}[t]
%\caption{Methods used for numerical evaluation.}
%\label{tab:evaluation_settings}
%\centering
%\setlength{\tabcolsep}{3pt}
%\renewcommand{\arraystretch}{0.95}
%\begin{tabularx}{\linewidth}{>{\raggedright\arraybackslash}p{0.23\linewidth}>{\raggedright%\arraybackslash}p{0.17\linewidth}>{\raggedright\arraybackslash}p{0.09\linewidth}Y}
%\toprule
%Method & Programmed validators & LLM use & Purpose \\
%\midrule
%Exact-match \prove{} & Yes & No & Non-LLM \prove{} variant using exact rule-label matching before programmed numerical validation. \\
%Fuzzy-match \prove{} & Yes & No & Non-LLM \prove{} variant using fuzzy lexical matching before programmed numerical validation. \\
%Embedding-match \prove{} & Yes & No & Non-LLM \prove{} variant using embedding similarity matching before programmed numerical validation. \\
%LLM-assisted \prove{} & Yes & Yes & LLM-assisted \prove{} variant, using language-model support for table interpretation and related language tasks while keeping numerical decisions in programmed validators. \\
%\bottomrule
%\end{tabularx}
%\end{table}

\section{Results}

\paragraph{Simulation data and discrepancy construction.}
Each simulation replicate evaluates one clean reporting package and one paired discrepancy-injected package generated from the same synthetic trial data. The injected discrepancies are selected from the rule classes described above so that the expected findings are known before validation. This structure allows performance to be assessed at both the package level and the rule level.

\paragraph{Simulation performance}
Table~\ref{tab:rule-set-a-full-tfl-performance} reports replication-wise performance for discrepancy-injected TFL packages under label variation. The comparison includes exact rule-label matching, fuzzy lexical matching, embedding similarity matching, LLM-assisted semantic matching, and manual review. For each metric and method, the table reports the mean across replications and the minimum, 25th percentile, median, 75th percentile, and maximum across replication-level results. Precision is calculated among reported findings, and recall is calculated among injected discrepancies.
{\revone Accordingly, recall in Table~\ref{tab:rule-set-a-full-tfl-performance} should be interpreted as within-library recall: the proportion of injected discrepancies from the implemented rule classes that were detected by the method.\locrone{R1.1.3}}
{\revone
The evaluation uses TFL output documents in Word/RTF-style formats rather than a pre-parsed table representation. The reported metrics reflect the implemented \prove{} pipeline starting from Word/RTF-style TFL output documents, including upstream ingestion and extraction/standardization steps, semantic label mapping, and programmed validation.
}

We evaluate findings using two matching definitions. The primary metrics require the finding to match the exact rule used to inject the discrepancy. The overall metrics give credit when the altered number is identified by any applicable rule, even if the reported rule differs from the rule used for injection. Thus, the overall metrics capture cases where a poisoned value is correctly recognized as problematic through a related check.

{\revone
%Exact-label results are reported in supplemental Table~\ref{tab:rule-set-a-full-tfl-performance-exact-label-supplement}.
The exact-label supplemental results show that all automated methods achieved perfect performance when table labels matched the validator vocabulary. This confirms that the programmed validators behave as expected when the relevant rows, columns, and statistics can be mapped unambiguously.

The results based on label variation are reported in  Table~\ref{tab:rule-set-a-full-tfl-performance}.
Under label variation, the LLM-assisted semantic matching variant had the strongest replication-wise performance across recall and F1, while the exact-match, fuzzy lexical, and embedding similarity baselines were more sensitive to terminology changes. 
The primary precision, recall, and F1 values for exact-match, fuzzy lexical, and embedding similarity matching were identical, indicating that neither fuzzy nor embedding matching recovered additional exact-rule matches beyond the exact-match baseline for the primary rule-level definition. Embedding similarity also matched the exact-match baseline for the overall metrics.  In this benchmark, embedding similarity did not improve over exact matching, possibly because many TFL labels are short clinical/statistical terms or abbreviations; this should be interpreted as an empirical result of the current settings rather than a general limitation of embedding methods.

%One likely explanation is that many of the relevant TFL labels are short clinical or statistical terms and abbreviations, for which the local general-purpose embedding model may not provide enough additional separation beyond exact normalized matching under the current threshold and margin settings. This result should therefore be interpreted as an empirical finding for this benchmark, not as evidence that embedding methods cannot help in other labeling regimes.

Fuzzy lexical matching provided a modest gain under the overall matching definition, increasing mean overall recall from 0.527 to 0.588 and mean overall F1 from 0.684 to 0.735 compared with exact matching, while maintaining high overall precision. However, this improvement remained substantially below the LLM-assisted semantic matching variant, which achieved mean recall of 0.993 and mean F1 of 0.996 with minimal cross-replication variation. Thus, the main benefit of LLM assistance in this experiment was improved robustness to label variation, rather than improved numerical validation, since all methods used the same programmed validators.

To further characterize missed detections under label variation, we summarized false negatives by injected rule class. The exact-match, fuzzy lexical, and embedding similarity variants showed the same dominant false-negative pattern. Missed findings were concentrated in rule classes that depend on recognizing statistic labels or cross-output relationships, including mean/range consistency (8 misses), cross-table related-count hierarchy (7 misses), cross-table count consistency (7 misses), Kaplan--Meier event/censoring consistency (6 misses), confidence-interval bound (6 misses), and absolute-zero p-value (6 misses). In contrast, the LLM-assisted variant had only one false negative, corresponding to a Kaplan--Meier accounting mismatch. Together with the perfect exact-label performance in these simulated replications, the false negatives observed under label variation are likely due to failures to associate rewritten labels, statistics, and related outputs with the corresponding validation checks, rather than failures of the programmed numerical validators.

Manual review showed lower and more variable detection performance than the automated methods under the same label-variation setting. The false positives were concentrated in clinically plausible but non-poisoned cells, especially PFS hazard-ratio, confidence-interval, percentile, and related summary rows, suggesting that reviewers often flagged neighboring or suspicious values that did not exactly match the injected discrepancy location. False negatives were common in tumor-response, systemic-therapy, subgroup PFS, DOR, and safety hierarchy outputs, where detecting the injected discrepancy often required recomputation, cross-row comparison, or cross-output reasoning under time constraints. Thus, the manual-review baseline reflects a realistic limited-time QC setting rather than an exhaustive independent-programming review.
}

\IfFileExists{generated/table_rule_set_a_full_tfl_performance.tex}{\begin{table}[t]
\centering
\setlength{\tabcolsep}{3pt}
\renewcommand{\arraystretch}{0.95}
\caption{Replication-wise poisoned full-TFL performance under label variation. Mean and percentile summaries are computed across replicate-level metric values.}
\label{tab:rule-set-a-full-tfl-performance}
\begin{tabular*}{\linewidth}{@{\extracolsep{\fill}}llrrrrrr}
\toprule
Metric & Method & Mean & Min & 25\% & Median & 75\% & Max \\
\midrule
\multirow{5}{*}{Precision} & Exact Match & 0.986 & 0.857 & 1.000 & 1.000 & 1.000 & 1.000 \\
 & Fuzzy Match & 0.986 & 0.857 & 1.000 & 1.000 & 1.000 & 1.000 \\
 & Embedding Match & 0.986 & 0.857 & 1.000 & 1.000 & 1.000 & 1.000 \\
 & LLM-assisted & 1.000 & 1.000 & 1.000 & 1.000 & 1.000 & 1.000 \\
 & Manual Review & 0.526 & 0.350 & 0.381 & 0.444 & 0.562 & 1.000 \\
\midrule
\multirow{5}{*}{Recall} & Exact Match & 0.474 & 0.385 & 0.429 & 0.481 & 0.500 & 0.583 \\
 & Fuzzy Match & 0.474 & 0.385 & 0.429 & 0.481 & 0.500 & 0.583 \\
 & Embedding Match & 0.474 & 0.385 & 0.429 & 0.481 & 0.500 & 0.583 \\
 & LLM-assisted & 0.993 & 0.929 & 1.000 & 1.000 & 1.000 & 1.000 \\
 & Manual Review & 0.467 & 0.200 & 0.385 & 0.500 & 0.567 & 0.667 \\
\midrule
\multirow{5}{*}{F1} & Exact Match & 0.638 & 0.556 & 0.600 & 0.633 & 0.667 & 0.737 \\
 & Fuzzy Match & 0.638 & 0.556 & 0.600 & 0.633 & 0.667 & 0.737 \\
 & Embedding Match & 0.638 & 0.556 & 0.600 & 0.633 & 0.667 & 0.737 \\
 & LLM-assisted & 0.996 & 0.963 & 1.000 & 1.000 & 1.000 & 1.000 \\
 & Manual Review & 0.442 & 0.333 & 0.412 & 0.474 & 0.481 & 0.500 \\
\midrule
\multirow{5}{*}{Overall precision} & Exact Match & 0.986 & 0.857 & 1.000 & 1.000 & 1.000 & 1.000 \\
 & Fuzzy Match & 0.988 & 0.875 & 1.000 & 1.000 & 1.000 & 1.000 \\
 & Embedding Match & 0.986 & 0.857 & 1.000 & 1.000 & 1.000 & 1.000 \\
 & LLM-assisted & 1.000 & 1.000 & 1.000 & 1.000 & 1.000 & 1.000 \\
 & Manual Review & 0.526 & 0.350 & 0.381 & 0.444 & 0.562 & 1.000 \\
\midrule
\multirow{5}{*}{Overall recall} & Exact Match & 0.527 & 0.385 & 0.471 & 0.536 & 0.571 & 0.667 \\
 & Fuzzy Match & 0.588 & 0.462 & 0.547 & 0.577 & 0.628 & 0.750 \\
 & Embedding Match & 0.527 & 0.385 & 0.471 & 0.536 & 0.571 & 0.667 \\
 & LLM-assisted & 0.993 & 0.929 & 1.000 & 1.000 & 1.000 & 1.000 \\
 & Manual Review & 0.467 & 0.200 & 0.385 & 0.500 & 0.567 & 0.667 \\
\midrule
\multirow{5}{*}{Overall F1} & Exact Match & 0.684 & 0.556 & 0.640 & 0.697 & 0.727 & 0.800 \\
 & Fuzzy Match & 0.735 & 0.632 & 0.707 & 0.732 & 0.771 & 0.857 \\
 & Embedding Match & 0.684 & 0.556 & 0.640 & 0.697 & 0.727 & 0.800 \\
 & LLM-assisted & 0.996 & 0.963 & 1.000 & 1.000 & 1.000 & 1.000 \\
 & Manual Review & 0.442 & 0.333 & 0.412 & 0.474 & 0.481 & 0.500 \\
\bottomrule
\end{tabular*}
\end{table}
}{\artifactplaceholder{Run the full-TFL simulation to generate paired clean/discrepancy-injected performance metrics.}}

%\paragraph{Representative findings.}
%\resultplaceholder{Add a concise example showing a detected percentage mismatch or safety hierarchy violation, including table ID, row, column, expected value, observed value, and source/evidence fields.}

\section{Discussion}

\prove{} demonstrates a trustworthy-AI pattern for statistical reporting workflows: LLMs may help interpret unstructured or semistructured materials, but programmed validators remain responsible for numerical and logical checks. This separation directly addresses risks that arise when probabilistic models are asked to perform arithmetic, denominator checks, or regulatory-style consistency verification without executable constraints.

The multi-method, two-label-setting simulation is important for both engineering and scientific interpretation. The exact-match, fuzzy lexical, and embedding similarity \prove{} variants measure performance when the full workflow relies on non-LLM label interpretation and programmed validation. The LLM-assisted \prove{} variant measures whether LLM-assisted interpretation improves detection in outputs whose labels or structure are less direct, while preserving programmed validation for numerical and logical decisions. Including fuzzy lexical and embedding similarity baselines helps distinguish the contribution of LLM-assisted semantic interpretation from simpler non-LLM paraphrase-matching approaches. The exact-label setting verifies that the automated variants behave as expected when the reporting package follows predefined terminology, while the label-variation setting tests whether the methods remain reliable under realistic wording changes.

The label-variation results also illustrate a practical role for LLMs in clinical reporting quality control. Across studies, TFLs often differ in row wording, column headers, endpoint labels, footnote style, and formatting conventions even when they report the same underlying clinical concepts. A purely programmed workflow can perform well when these labels are prespecified, but it may miss checks when equivalent concepts are expressed using different language. In \prove{}, LLM support improves robustness by helping map varied table language back to the concepts expected by the validator, while the final numerical assessment remains governed by executable checks. This division is important because it uses the LLM where flexibility is needed and preserves deterministic behavior where correctness must be verifiable.

\paragraph{Limitations}
{\revone The current simulation uses synthetic data and controlled discrepancy injection to support reproducible rule-level evaluation; therefore, recall should be interpreted as conditional recall within the implemented QC rule library rather than over the natural distribution of clinical reporting discrepancies, which would require external evidence such as real QC findings, coded historical errata or independently generated discrepancies. At the current stage, PROVE is tested for oncology trial TLF outputs. The exact-label results indicate that extraction and standardization were well controlled for the current synthetic Word-format TFL benchmark, but upstream document handling may require format-specific adaptation and validation for sponsor-specific Word/RTF formats, unusual table layouts, or heavily footnote-driven outputs. General validation patterns for counts, percentages, denominators, descriptive statistics, time-to-event summaries, and safety outputs are expected to transfer across therapeutic areas, whereas oncology-specific endpoint rules and semantic labels would need adaptation for cardiovascular, hematology, rare disease, or other therapeutic areas. Future work will expand the rule library using historical QC logs, independently generated review findings, and published regulatory errata, and will adapt PROVE to handle real sponsor packages with non-standard table layouts, merged cells, sponsor-specific terminology, and footnote-driven definitions that affect denominator, population, or endpoint interpretation. 
Results in the simulations should not be interpreted as evidence that upstream document handling would be error-free for all sponsor formats. 
A more direct evaluation of row-concept, denominator, and rule-mapping precision/recall would require generator-exported or independently annotated gold-standard mappings for intermediate table representations; this will be considered in future benchmark extensions.}

%\paragraph{Reproducibility.}
%We provide a public Streamlit application at \url{https://provetlf2026.streamlit.app/}. The application exposes the same simulation settings for \prove{} with exact-match, fuzzy lexical, embedding similarity, or LLM-assisted semantic matching support.

\section{Disclaimer}
Please note the views and opinions expressed in this presentation are those of the authors and are not intended to reflect the views and/or opinions of their employers.

\section{Conclusion}
\prove{} provides an auditable, role-separated framework for clinical trial output quality control by combining controlled language-model support for interpretation with programmed statistical validation for numerical and logical decisions. This design supports trustworthy AI evaluation by preserving source provenance for each finding and measuring performance through a reproducible simulation benchmark. By separating semantic interpretation from mathematical verification, \prove{} can assist clinical reporting review while keeping final correctness decisions in executable validation rules.
%\prove{} provides an auditable, agentic framework for clinical trial outputs quality control by integrating deterministic statistical validation with LLM support. This architecture is uniquely suited for trustworthy AI evaluation because it maintains rigorous source provenance for every finding and quantifies performance through simulation. By providing a scalable and reproducible methodology, the framework enhances the precision and efficiency of regulatory submissions and strategic decision-making. Ultimately, the separation of semantic interpretation from mathematical verification ensures that \prove{} serves as a reliable assistant in high-stakes drug development environments.

%\begin{ack}
%Acknowledgments are hidden in anonymous submission mode and will be added for the final version.
%\end{ack}

\bibliographystyle{plainnat}
\bibliography{reference}

\end{document}